\documentclass[%
 reprint,
 amsmath,amssymb,
 aps,
prb,
]{revtex4-1}

\usepackage{graphicx}% Include figure files
\usepackage{dcolumn}% Align table columns on decimal point
\usepackage{bm}% bold math
\begin{document}

\preprint{APS/123-QED}

\title{Magnonic band gaps in waveguides with a periodic variation of the saturation magnetization}% Force line breaks with \\

\author{Florin Ciubotaru}
  \email{ciubotaru@physik.uni-kl.de}
\author{Andrii~V.~Chumak}
\author{Bj\"orn~Obry}
\author{Alexander A. Serga}%
\author{Burkard Hillebrands}
\affiliation{%
 Fachbereich Physik and Landesforschungszentrum OPTIMAS, Technische Universit\"{a}t Kaiserslautern, 67663 Kaiserslautern,
 Germany}%

\date{\today}% It is always \today, today,

\begin{abstract}
We present a micromagnetic analysis of spin-wave propagation in a
magnonic crystal realized as a permalloy spin-wave waveguide with
a spatial periodical variation of its saturation magnetization.
Frequency band gaps were clearly observed in the spin-wave
transmission spectra and their origin is traced back to an overlap
of individual band gaps of the fundamental and the higher order
spin-wave width modes. The control of the depth, width and the
position in frequency and space of the rejection band gaps by the
width areas with a reduced magnetization and by the modulation
level, are discussed in this study.

\begin{description}
\item[PACS numbers]{75.30.Ds, 75.78.Cd, 85.70.Kh}
\end{description}
\end{abstract}

\maketitle

%\section{INTRODUCTION}

The transmission spin-wave spectra in magnetic materials with a
periodic variation of their properties - magnonic crystals (MC) -
show evidence of special properties that are distinctive from
those of uniform media. Most noticeable is the appearance of
rejection bands, i.e. frequency intervals over which the
propagation of spin waves is forbidden.\cite{YuGulyaev,SVysotskii}
Furthermore, the spin-wave group velocity and phase are strongly
modified at the band gap edges and they can be artificially
manipulated.\cite{ChumakPRL} All these properties make magnonic
crystals good candidates for signal processing and information
transfer devices such as delay lines,\cite{Ustinov}
logic\cite{Ding} and storage elements.\cite{ChumakPRL}

A variety of magnonic crystal designs have been proposed and
tested in order to improve their operational characteristics and
to realize new signal processing procedures. For example, magnonic
crystals are fabricated from different magnetic
materials,\cite{Wan,Kruglyak} with different
shapes\cite{KimHan,ChumakmuMC,Ciubotaru} or with a local variation
of the bias field. \cite{ChumakCurrent,ChumakReversal} The best
transmission properties were achieved for MCs that are based on
yttrium-iron-garnet (YIG) ferrite films, due to the extremely
small spin-wave losses in this material.
\cite{APL-chumak,APLChumak} However, modern microwave devices
require micro-sized elements which can hardly be produced from
YIG. Previous studies performed on shape-modulated permalloy
waveguides have shown good transmission characteristics with deep
band gaps. Moreover, the possibility to control the number of the
band gaps by a proper choice of the MC geometry was
demonstrated.\cite{ChumakmuMC,Ciubotaru} Nevertheless, these MCs
are characterized by a non-resonant spurious scattering on each
geometric non-uniformity of the waveguide which results in an
increase of the transmission loss and a broadening of the
frequency gap. This effect is not always desirable in the signal
processing.\cite{Ciubotaru,KimHan} A reliable substitute for a
shape modulated crystal is given by a bi-component MC where two
magnetic materials are placed
periodically.\cite{KrawczykPRB,Wan,Gubbiotti,Kostylev} However,
fabrication of a two-component MC involves multi step and high
precision e-beam lithography and film deposition which complicates
the fabrication procedure. An alternative design comprises
structures with a periodical variation in the saturation
magnetization by an ion implantation technique.\cite{Obry}

Here we present a systematic micromagnetic study of a
one-component magnonic crystal based on a permalloy (Py) waveguide
with a periodic variation of its saturation magnetization. Such
structures can be realized, e.g., by local ion
irradiation.\cite{Obry} We show that the spin-wave transmission
characteristics exhibits pronounced frequency band gaps which
depend on both the level of the magnetization modulation and the
size of the areas with a reduced magnetization. Furthermore, the
role of the higher-order spin-wave width modes characterized by a
quantization number of the wavevector component transverse to the
stripe due to the finite width, is also discussed.

\begin{figure}[b]
    \includegraphics[width=1\columnwidth]{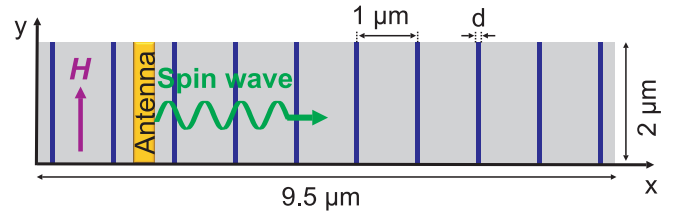}
    \caption{\label{FCFig1} (Color online) Sketch
of the simulated structures: MC based on a Py waveguide with a
periodic variation of its saturation magnetization. The dark blue
lines denote the areas where the saturation magnetization has been
reduced by a certain percentage. }
\end{figure}

\begin{figure*}[t]
    \includegraphics[width=2\columnwidth]{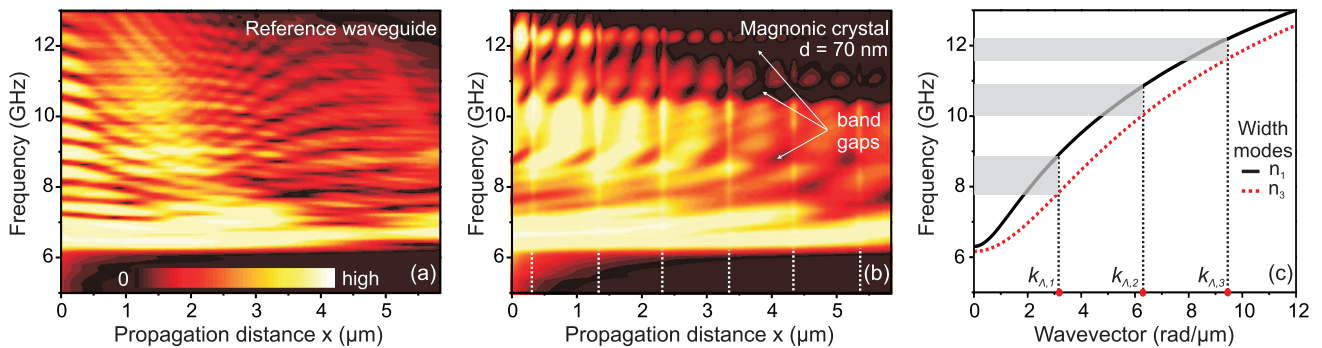}
    \caption{\label{FCFig2} (Color online) Spin-wave transmission characteristics for a simple permalloy
waveguide (a) and a magnonic crystal (b) as a function of the
propagation distance ($x$) from the excitation antenna. The dashed
white lines in b) denote the positions of the areas where the
saturation magnetization was decreased. (c) Dispersion relations
calculated for the first and the third spin-wave width modes. The
shaded regions mark the frequency intervals corresponding to the
band gaps observed in the transmission characteristic in (b),
while $k_{\Lambda,n}$ represent the Bragg reflection wavenumbers.}
\end{figure*}

The magnonic crystal structure is presented in Fig.~\ref{FCFig1}.
A 40 nm thick permalloy waveguide with a width of 2~$\mu$m is
magnetized transverse to its long axis by a static biasing
magnetic field of $B_{0}=50$~mT. The length of the waveguides is
9.5 $\mu$m.  The spatial variation of the magnetization was
expressed according to a step-like function. The MC lattice
constant, i.e. the periodicity of the area where the saturation
magnetization was decreased, is $\Lambda=1$~$\mu$m. The
simulated\cite{OOMMF} area was discretized into $N_{x}\times
N_{y}\times N_{z}=950\times200\times4$ cells, each cell having a
size of $10\times10\times10$~nm$^{3}$. The standard material
parameters of Py were used: saturation magnetization
$\mu_{0}M_{0}=1$~T, exchange stiffness constant $A =
1.3\times10^{-11}$~J/m and zero magnetocrystalline anisotropy. In
order to avoid spin-wave reflection at the ends of the structure
($x = 0$ and $x = 9.5$ $\mu$m) the following damping boundary
conditions were used: in the boundary areas ($\sim1\mu$m on each
side) the damping parameter $\alpha$ gradually increases more than
fifty times resulting in a strong decay of the spin-wave intensity
towards the boundaries \cite{Dvornik1}.

In order to excite spin waves within a wide frequency range we
apply a rectangular magnetic field pulse with a duration of 70~ps
and an amplitude of 3~mT. The field is oriented in the
$x$-direction and is uniformly distributed in a 300~nm wide area
across the Py waveguide. This area plays the role of a spin-wave
excitation antenna. The cut-off frequency limit given by the pulse
duration is $\sim$13.5~GHz while the antenna can excite spin-waves
with wavenumbers up to $k_\mathrm{max} = 21$~rad/$\mu$m (which
corresponds to a frequency of 54.6~GHz at $B_{0}=50$~mT applied
field). The applied static magnetic field satisfies the conditions
for the excitation and propagation of magnetostatic surface spin
waves (MSSWs). These waves are known to have the highest values of
the group velocity in magnetic microstructures and ensure a
propagation distance in the order of several micrometers in a
permalloy film.

As a reference, we simulated a regular permalloy waveguide with
the same dimensions as the magnonic crystal. The spin-wave
intensity extracted from the simulation as a function of the
frequency and of the propagation distance from the antenna is
displayed in Fig.~\ref{FCFig2}(a). One can observe that no
spin-wave transmission is allowed below a frequency of 6.1~GHz,
which corresponds to the ferromagnetic resonance (FMR) frequency,
i.e. spin waves with zero wavenumbers. The non-zero intensities
observed below $\sim$6.1~GHz are a result of the forced excitation
of the magnetization dynamics and are only detected close to the
antenna. The maximum intensity corresponds to long-wavelength spin
waves with the highest excitation efficiency and the highest group
velocity that are excited just above the FMR frequency. With
increasing frequency and, implicitly, with increasing spin-wave
wavenumber, the excitation efficiency decays.\cite{Demidov1}
Furthermore, one can notice a frequency and a distance dependent
oscillation of the spin-wave intensity even for the case of a
regular Py waveguide. These fluctuations suggest that interference
between the fundamental and higher order width modes play an
important role in the spin-wave propagation.\cite{Buttner,PPirro}

\begin{figure*}[t]
    \includegraphics[width=1.6\columnwidth]{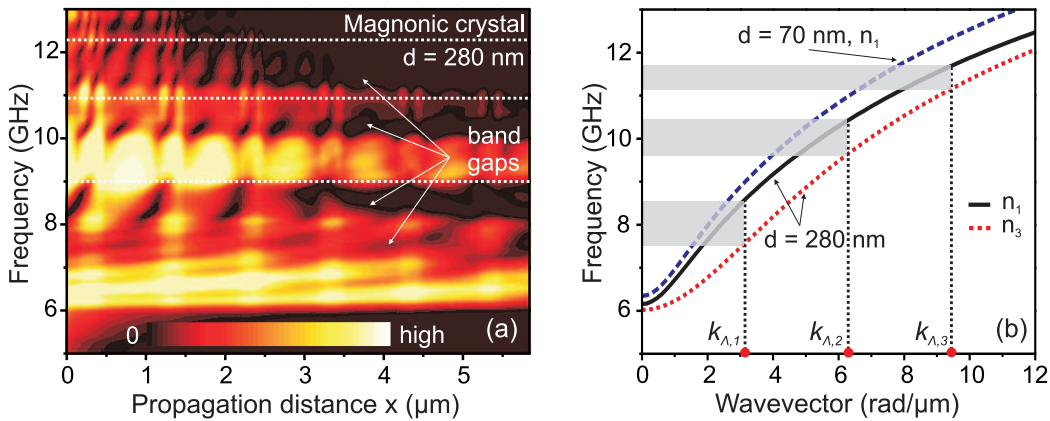}
    \caption{\label{FCFig3} (Color online) (a)
Spin-wave transmission characteristics as a function of the
propagation distance (x) from the excitation antenna for a
magnonic crystal with the ``doping'' area length $d=280$~nm. The
dotted lines represent the frequency band gap position for a MC
with $d=70$~nm (Fig.~\ref{FCFig2}b). (b) Dispersion relations
calculated taking into account a saturation magnetization averaged
over the entire crystal for a MC with $d=70$~nm (first width mode)
and a MC with $d=280$~nm (first and third width modes). The shaded
regions mark the frequency intervals corresponding to the band
gaps observed in the transmission characteristic in (a),
$k_{\Lambda,n}$ represent the Bragg reflection wavenumbers. }
\end{figure*}

The spin-wave intensity map obtained by simulating a magnonic
crystal with a periodical reduction of 10\% of the saturation
magnetization over an area width $d=70$~nm remains virtually
unchanged (figure not shown). However, a decrease of 20\% in the
saturation magnetization changes the spin-wave intensity map
drastically, as can be seen in Fig.~\ref{FCFig2}(b). Three
spin-wave band gaps are visible and their frequencies are closely
correlated with the Bragg reflection wavenumbers as can be seen by
looking at the dispersion relations displayed in
Fig.~\ref{FCFig2}(c). The spin-wave band gap corresponding to the
first Bragg wavenumber $k_{\Lambda1} = \pi/\Lambda =
3.14$~rad/$\mu$m is visible at $\sim$8~GHz. The second band gap
($k_{\Lambda2} = 2\pi/\Lambda = 6.28$~rad/$\mu$m) and the third
one ($k_{\Lambda3} = 3\pi/\Lambda = 9.42$~rad/$\mu$m) are very
pronounced near 10.7~GHz and 11.8~GHz, respectively. The fact that
the first band gap is not as clearly defined can be understood by
taking into account that the antenna excites higher order width
spin-wave modes along with the fundamental one. The Bragg
reflection for each width mode occurs at different frequencies
(see Fig.~\ref{FCFig2}(c)). At a given frequency, when one mode is
Bragg reflected, the other width modes are still transmitted, thus
the total intensity is not as strongly affected. However, this
effect becomes less important for higher-order band gaps since the
frequencies of the width modes are getting closer to each other.
Thus, one can conclude that the higher order spin-wave width modes
play an important role: The frequency position and width of the
band gaps are determined by the superposition of the partial band
gaps formed by individual spin-waves width modes. It is worth
mentioning that the wavelength of spin waves is not only
determined by the local magnetization but is sensitive to the
average magnetization of the entire sample. Therefore, the spin
wave dispersion relations from Fig.~\ref{FCFig2}(c) were
analytically calculated taking into account an average saturation
magnetization of the structure of
$\mu_{0}M_{av}=\mu_{0}[M_{0}-\frac{da}{L}(M_{0}-M)]=0.985$~T where
$a$ is the number of regions with reduced magnetization and L is
the total length of the waveguide.

Comparing the two spin-wave intensity maps obtained for a regular
Py waveguide and for the 20\% MC (Fig.~\ref{FCFig2}) one can
remark that the decay of the spin-wave amplitude over distance is
the same for both structures. In addition, a periodic variation of
the spin-wave intensity as a function of the propagation distance
is clearly visible inside of MC. This variation manifests in the
appearance of local intensity maxima in the crystal areas with
unchanged saturation magnetization (each single period). One
maximum at 10~GHz, two maxima at 11~GHz and three at 12.3~GHz are
visible inside these areas. They can be understood as a formation
of quasi-standing modes due to multiple reflections at the
boundaries where $M_{0}$ was reduced. It should be noted that the
wavelength of the quasi-standing modes fulfill the condition
$\lambda\approx na/2$, where $n$ is an integer parameter and $a$
is the width of the area of unchanged magnetization. These
observations underline the fact that a magnonic crystal works as a
series of coupled resonators.

To control the frequency of the band gaps it should be taken into
account that they are mainly determined by the Bragg wavenumbers
and the spin-wave dispersion relation. The dispersion relation is
strongly dependent on $M_{0}$ and on the bias magnetic field. A
change in the magnetic field or in the saturation magnetization
shifts the spin-wave dispersion relation up or down and,
implicitly, increases or decreases the frequency of the band gap
at the fixed Bragg wavenumbers. A variation of the MC lattice
constant $\Lambda$ changes the wavelength of spin waves that
satisfy the reflection Bragg condition $2\Lambda=n\lambda$ ($n$
integer), and consequently the band gap frequency. However, the
above parameters do not provide any information about the depth of
the band gaps. To have a complete overview of the band gaps
behavior we performed a systematic analysis by manipulating both
the width of the implanted areas and the level of the
magnetization variation $M/M_{0}$ of the MC structure.

By simulating an MC with a 30\% reduction with respect to $M_{0}$
we observed a stronger attenuation even for the transmitted spin
waves. For example, above 10~GHz the spin waves can propagate for
only $\sim$1.5~$\mu$m. This decay is caused by an increased
reflection at the boundaries between areas with different
magnetization values. Keeping a 20\% reduction of $M_{0}$ and
increasing the length of the doping area by increasing $d$ from
70~nm to 280~nm we observe deeper band gaps while the propagation
distance in the transmission bands remains nearly unchanged, as
can be seen in Fig.~\ref{FCFig3}(a). It can be concluded that a
band gap depth control can be achieved by manipulating the width
of the areas with a reduced magnetization. Furthermore, the band
gaps for the latter structure are shifted to lower frequencies
(the white dotted lines mark the position of the band gaps for the
MC with $d=70$~nm). This fact can be understood by analyzing the
dispersion relations (see Fig.~\ref{FCFig3}(b)) calculated using
an average saturation magnetization of the entire structure of
$\mu_{0}M_{av}=0.941$~T. Considering this, one can observe that
the band gap positions obtained from the simulation match very
well the ones calculated analytically. In addition, from
Fig.~\ref{FCFig3}(a) one can observe the formation of a
supplementary band gap at a frequency of 7.5 GHz, which
corresponds to a Bragg reflection of the third spin-wave width
mode. This effect is visible at lower frequencies where there is a
large frequency interval between the first and the third width
modes at the first Bragg wavenumber $k_{\Lambda,1}$ (see
Fig.~\ref{FCFig3}(b)). Furthermore, one should note that the band
gaps widths are not changed by using larger areas with a decreased
magnetization.

\begin{figure}[t]
    \includegraphics[width=1.0\columnwidth]{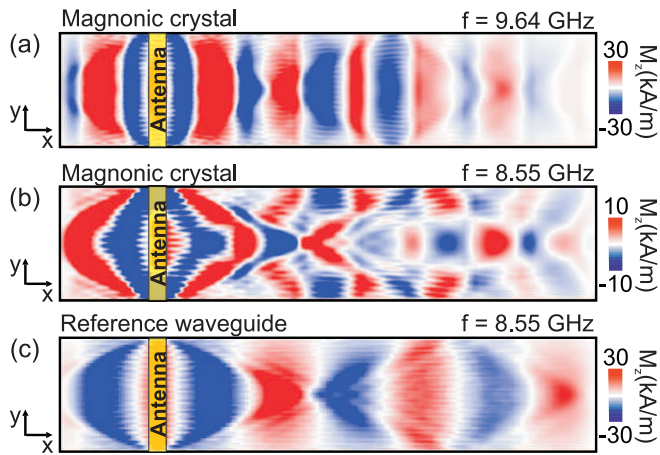}
    \caption{\label{FCFig4} (Color online) Snapshot images of the magnetization oscillation pattern
with the component $M_{z}$ color coded recorded for the following
cases: MC for a continuous excitation with a frequency of 9.64 GHz
(a) and 8.55 GHz (b); (c) reference waveguide for an excitation
frequency of 8.55 GHz. }
\end{figure}

Each spin-wave width mode possesses individual band gaps for the
wavenumbers that satisfy the Bragg condition. Therefore, exciting
exactly with a frequency for which only one width mode is
reflected the other modes will be transmitted. For example, using
an excitation frequency of 9.64~GHz (second band gap at
$k_{\Lambda,2}$ for the third width mode, see
Fig.~\ref{FCFig3}(b)) the magnetization oscillation pattern shows
a propagation of only the first width mode, as can be seen in
Fig.~\ref{FCFig4}(a). If a frequency of 8.55~GHz is used (band gap
at $k_{\Lambda,1}$ for the first width mode) the magnetization
oscillation patterns evidence the fact that only the third width
mode is transmitted (Fig.~\ref{FCFig4}(b)). The two spatial
distributions of the magnetization are compared with the one
obtained for the reference waveguide (Fig.~\ref{FCFig4}(c)) where
a typical interference pattern\cite{Buttner,Demidov2} between the
first and the third width modes is observed. An excitation
frequency of 8.55~GHz was used for the latter case. Since
different spin-waves width modes can be suppressed at chosen
excitation frequencies makes the magnonic crystal to act as a mode
selective filter.

In summary, we performed a micromagnetic study of the spin-wave
transmission in 1D magnonic crystals based on a permalloy
waveguide with a periodic variation of its saturation
magnetization. The appearance of frequency band gaps was clearly
observed and studied in the space and frequency domains. The band
gaps are determined by the superposition of individual band gaps
of the fundamental and of the higher order spin-wave width modes.
Furthermore, the rejection bands depend strongly on both the level
of the magnetization variation and the size of the area with a
reduced magnetization: The reduction of the saturation
magnetization over a larger area leads to the formation of more
pronounced band gaps. At the same time the magnetization $M_{av}$
averaged over the MC must be taken into account for the spin-wave
dispersion relation: $M_{av}$ shifts the spin wave dispersion
characteristics down, i.e. the band gaps appear at lower
frequencies.

\section*{ACKNOWLEDGMENTS}

Financial support by the DFG SE-1771/1-2 is gratefully
acknowledged.

\end{document}